# X-Band ESR Determination of Dzyaloshinsky-Moriya Interaction in 2D SrCu$_2$(BO$_3$)$_2$ System

Andrej Zorko,[1,*] Denis Arčon,[1] Hans van Tol,[2] Louis Claude Brunel,[2] and Hiroshi Kageyama[3]

[1] *Institute "Jožef Stefan", Jamova 39, 1000 Ljubljana, Slovenia*

[2] *National High Magnetic Field Laboratory, Florida State University, Tallahassee, FL 32310, USA*

[3] *Department of Chemistry, Graduate School of Science, Kyoto University, Kyoto, 606-8502, Japan*

X-band ESR measurements on a single crystal of SrCu$_2$(BO$_3$)$_2$ system in a temperature range between 10 K and 580 K are presented. The temperature and angular dependence of unusually broad ESR spectra can be explained by the inclusion of antisymmetric Dzyaloshinsky-Moriya (DM) interaction, which yields by far the largest contribution to the linewidth. However, the well-accepted picture of only out-of-plane interdimer DM vectors is not sufficient for explanation of the observed angular dependence. In order to account for the experimental linewidth anisotropy we had to include sizable in-plane components of interdimer as well as intradimer DM interaction in addition to the out-of-plane interdimer one. The nearest-neighbor DM vectors $\mathbf{D}_\perp$ lie perpendicular to crystal anisotropy **c**-axis due to crystal symmetry. We also emphasize that above the structural phase transition occurring at 395 K dynamical mechanism should be present allowing for instantaneous DM interactions. Moreover, the linewidth at an arbitrary temperature can be divided into two contributions; namely, the first part arising from spin dynamics governed by the spin Hamiltonian of the system and the second part due to significant spin-phonon coupling. The nature of the latter mechanism is attributed to phonon-modulation of the antisymmetric interaction, which is responsible for the observed linear increase of the linewidth at high temperatures.

[*] Corresponding author: tel.: +386-1-4773866; fax: +386-1-4263296; e-mail: andrej.zorko@ijs.si

# I. Introduction

Low-dimensional quantum spin systems have been intensively studied over past years, both from the experimental as well as the theoretical point of view. Quantum spin fluctuations often play a crucial role in determining the ground state and low-lying magnetic excitations of these systems. In this manner a nonmagnetic singlet ground state with a spin gap to the first magnetically excited state is a result of competing exchange interactions or significant frustration present in the system. Such a spin gap can be found in many one-dimensional (1D) spin systems but only in few two-dimensional (2D) systems.

Recently, a new 2D spin-gap system $SrCu_2(BO_3)_2$ was discovered with a special orthogonal network of $Cu^{2+}$ dimers formed out of localized $S = 1/2$ spins (see Fig. 1).[1] Various magnetic properties of this system are well described by the Hamiltonian of the so called 2D orthogonal dimer model,[2] which takes into account antiferromagnetic exchange interaction $J$ of each spin with its nearest neighbor (*nn*) as well as antiferromagnetic exchange coupling $J'$ to four next-nearest neighbors (*nnn*), i.e. $H_{ex} = J\sum_{(ij)} \mathbf{S}_i \cdot \mathbf{S}_j + J'\sum_{\{lm\}} \mathbf{S}_l \cdot \mathbf{S}_m$, where the sums run over all pairs of spins.

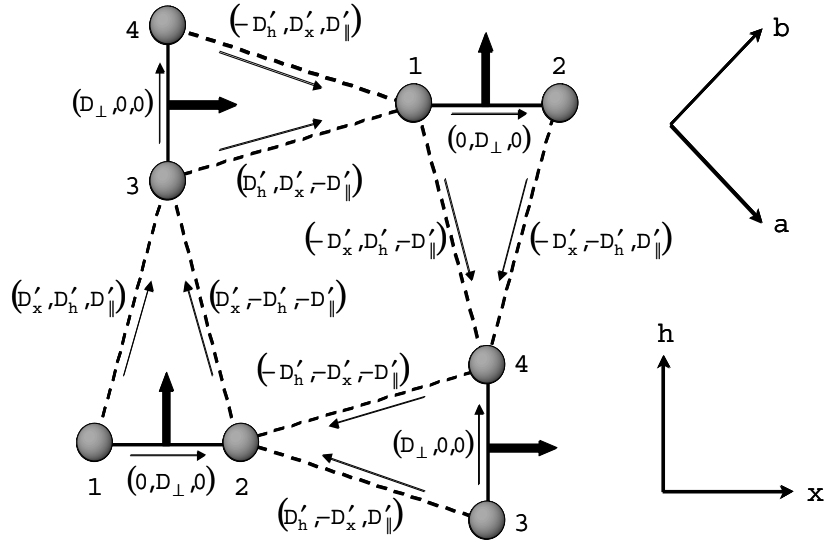

**Fig. 1:** The 2D network of $Cu^{2+}$ ions with full lines representing the *nn* ($J$) and dashed lines the *nnn* ($J'$) exchange coupling. The interdimer D and intradimer D' DM interactions are also presented with the thin arrows representing the orientation of corresponding bonds for DM interaction. Thick arrows are the *nn* DM vectors.

The 2D model of $SrCu_2(BO_3)_2$ spin system is topologically equivalent to the model considered by Shastry and Sutherland more than twenty years ago with an exactly solvable ground state.[3] Namely, the ground state is simply a product of singlets on each dimer up to a critical ratio of exchange constants $(J'/J)_c = 0.68$.[4] This singlet state remains the ground state of the system even when interdimer exchange $J''$ is taken into account.[5] Various sets of exchange parameters have been reported. Initially, $J = 100$ K and $J' = 0.68J$ was obtained based on the temperature dependence of the susceptibility.[2] Knetter et al. suggested $J = 71$ K, $J' = 0.6J$ and $J'' = 0.21J$ from the bound state energies of the two-triplet excitations.[6] However, recently the most frequently used set of exchange parameters is $J = 85$ K, $J' = 0.63J$ and $J'' = 0.09J$, which reproduces the temperature dependence of the magnetic susceptibility most accurately.[7] It should be stressed that the experimentally determined values of isotropic exchange constants place $SrCu_2(BO_3)_2$ compound to the extreme vicinity of the phase boundary $(J'/J)_c$ between the dimer phase and antiferromagnetically ordered phase.

Within the orthogonal dimer model the lowest-lying magnetically excited state is a single triplet excitation present on one of the dimers with the spin gap significantly suppressed with respect to the intradimer exchange $J$.[2] Experimentally, the first direct observation of the spin gap $\Delta = 35$ K was made by high-field ESR experiments[8] and inelastic neutron scattering measurements.[9] Since then its existence has been confirmed by various spectroscopic methods. Relatively small value of the spin gap as well as extremely localized nature[9] of a single triplet excitation are both due to a strong geometrical frustration of the *nnn* exchange interaction $J'$. On the other hand, the anisotropic behavior of the single triplet excitation strongly depending on the direction of applied external magnetic field cannot be explained with the oversimplified Hamiltonian $H_{ex}$ including only isotropic exchange interactions. To explain the fine structure of the triplet excitation Cépas et al.[10] included antisymmetric Dzyaloshinsky-Moriya (DM) exchange interaction into the spin Hamiltonian. Based on the symmetry arguments[11] the proposed form $H_{DM} = \sum_{\{ij\}} \pm D'_{\parallel} \mathbf{e}_c \cdot \mathbf{S}_i \times \mathbf{S}_j$ takes into account only *nnn* interactions with the interdimer Dzyaloshinsky-Moriya vector of a magnitude $D'_{\parallel} = 2.1$ K pointing in the crystal **c**-direction, i.e. parallel to the crystal anisotropy axis, which is perpendicular to 2D $Cu^{2+}$ plane (Fig. 1). Namely, neglecting a small buckling of the $CuBO_3$ plane,[12] this plane can be considered as a mirror plane, which assigns the direction of DM vectors parallel to the anisotropy axis. In addition, the intradimer DM interaction should vanish due to the center of inversion positioned at the middle of each dimer bond. Though this interaction reproduces the observed splitting of the single triplet excitation quite nicely and is of the expected amplitude $D'_{\parallel} \approx \Delta g/g \cdot J' \approx 6$ K, it cannot satisfactory explain the observation of singlet-triplet transitions in the high-field ESR experiments[8]. These transitions are in principle allowed since DM interaction mixes a finite amount of excited states into the ground singlet state. However, due to the symmetry of the system, interdimer DM interaction with DM vectors perpendicular to *ab*-plane still yields vanishing matrix elements between the ground state and the excited states.[13] The magnetic anisotropy of $SrCu_2(BO_3)_2$ is thus at the moment rather unclear and urges for new experiments.

In order to determine the major magnetic anisotropy interactions in $SrCu_2(BO_3)_2$ we performed a detailed single crystal X-band electron spin resonance (ESR) study in the temperature range between 10 K and 580 K, focusing in particular on the temperature and angular dependence of the ESR linewidth. ESR is in this respect a particularly powerful spectroscopic technique as the ESR absorption spectra are intimately related to spin anisotropy present in the system. Magnetic anisotropy manifests itself in an enhancement of linewidths and lineshifts of observed absorption lines. Temperature dependence of the resonance spectra reflects spin dynamics governed by anisotropic interactions of the investigated system.

Furthermore, conducting the experiment on the single crystal can potentially help to determine the direction of anisotropy axes of different contributions and their relative values through the angular dependence of parameters obtained from absorption spectra. X-band ESR study can thus yield further information on magnetic anisotropy of the system, which could not yet been evaluated or even observed by other spectroscopic techniques.

## II. Experimental

Experiments were performed on a single crystal of $SrCu_2(BO_3)_2$ compound of a size 5ä2ä2 mm$^3$, which was grown by the traveling solvent floating zone (TSFZ) method.[14] Its purity was verified by bulk susceptibility measurements down to 2K while the orientation of crystal axes was determined from Laue X-ray back-reflection. X-band ESR measurements were performed on the commercial Bruker E580 FT/CW spectrometer at the Larmor frequency of $\nu_L \sim 9.5$ GHz. The measurements in the temperature range between 10 K and the room temperature were conducted using an Oxford-cryogenics liquid-helium-flow cryostat while measurements at temperatures between the room temperature and 580 K required a use of a high-temperature controller with preheated-nitrogen-flow cryostat. High-field ESR spectra on powdered sample of $SrCu_2(BO_3)_2$ at the frequency of $\nu_L = 93.2$ GHz were recorded on a home build spectrometer working in a transition mode. The same high-purity polycrystalline sample was used for this measurement as in our previous X-band ESR report.[15]

## III. Theoretical background

In the case of a large isotropic exchange interaction one can divide the Hamiltonian of a spin system in an external magnetic field into two parts $H = H_0 + H'$, where $H_0 = H_Z + H_{ex}$ represents a sum of Zeeman interaction $H_Z$ and isotropic exchange coupling $H_{ex}$. All the spin anisotropy terms are contained in $H'$ and can be treated as a perturbation to the $H_0$ part of the Hamiltonian. In general, at temperatures well above the characteristic Zeeman splitting ($k_B T$ à $g\mu_B B_0$) the ESR absorption profile is formally given by the Fourier transform of the relaxation function $\varphi(t)$ as[16]

$$I(\omega - \omega_0) = \int_{-\infty}^{\infty} \varphi(t) \exp[i(\omega - \omega_0)t] dt, \qquad (1)$$

where the relaxation function $\varphi(t) = \langle \tilde{M}^+(t) M^-(0) \rangle / \langle M^+ M^- \rangle$ reflects fluctuations of transverse magnetization operator $M^+ = g\mu_B \sum_i S_i^+$ averaged over ensemble and $\tilde{M}$ denotes interaction representation. At high temperatures ($T$ à $J$) Kubo-Tomita approach[16] to magnetic resonance is justified and well established. In the case of Gaussian random processes the relaxation function is determined by the spin correlation function $\psi(\tau) = k_B^2 \langle [\tilde{H}'(\tau), S^+(0)][S^-(0), \tilde{H}'(0)] \rangle / \hbar^2 \langle S^+ S^- \rangle$ as

$$\varphi(t) = \exp\left[-\int_0^\infty (t-\tau)\psi(\tau)d\tau\right]. \tag{2}$$

Further simplifications are possible when spin-diffusional behavior of the spin correlation function, which is likely to set-in at high temperatures, is not effective and additionally the isotropic exchange interaction, which determines typical correlation time $\tau_c \sim \hbar/k_B J$ of the decay of the relaxation function, is large compared to the observed ESR peak-to-peak linewidth $\Delta B_{pp}$ ($k_B J \gg g\mu_B \Delta B_{pp}$). In this case the Fourier transform of the relaxation function $\varphi(t) = \exp[-t\psi(0)\tau_c]$ yields an absorption spectrum with a Lorentzian profile and the peak-to-peak linewidth

$$\Delta B_{pp} \approx \frac{2}{\sqrt{3}} \frac{k_B^2}{\hbar g\mu_B} M_2 \tau_c \approx \frac{2}{\sqrt{3}} \frac{k_B}{g\mu_B} \frac{M_2}{J}, \tag{3}$$

where the second moment of the resonance spectrum is given by the relaxation function at zero time, i.e. $M_2 = -\langle[H', S^+][S^-, H']\rangle / \langle S^+ S^-\rangle$. However, in low-dimensional systems a slow (potential) decay of spin correlation function at long times due to spin diffusion is often present at higher temperatures. It is determined by the dimensionality $d$ of the system as $\psi(\tau) \propto \tau^{-d/2}$ and gives rise to divergence of the correlation time $\tau_c$.[17] In this case significant deviations from Lorentzian profiles and additional line broadening are observed. A much better approximation for the correlation time $\tau_c$ is obtained by considering also the forth moment of the resonance line $M_4 = \langle[H - H_Z, [H', S^+]][H - H_Z, [H', S^-]]\rangle / \langle S^+ S^-\rangle$ yielding the linewidth of the following form[18]

$$\Delta B_{pp} \approx C \frac{k_B^2}{g\mu_B} \left(\frac{M_2^3}{M_4}\right)^{1/2}. \tag{4}$$

The constant $C$ appearing in Eq. (4), however, is not rigorously determined. Since the second and the forth moment of a pure Lorentzian line diverge, one is forced to carry out further approximations. The usually considered cut-off Lorentzian with a cut-off field at an arbitrary position yields a constant $C = \pi/3$. On the other hand, a more realistic lineshape given by the Lorentzian function multiplied by the exponential function of the form[18] $\exp[-(B - B_0)g\mu_B/k_B J]$ produces a constant of $C = 2\pi/\sqrt{6}$. It has to be emphasized that the exchange coupling in $SrCu_2(BO_3)_2$ system is few orders of magnitude larger than the corresponding external magnetic field in X-band ESR experiments, which means that no deviations from Lorentzian lineshape are expected to be observed despite the multiplication with the exponential function.

In low-dimensional spin systems with significant g-shifts and large isotropic exchange coupling the antisymmetric Dzyaloshinsky-Moriya interaction is often the dominant spin anisotropy contribution. However, it can be considerably suppressed by the symmetry of the system.[11] In the case when the DM interaction is the largest anisotropic interaction, the second and the forth moment of the absorption line are given in the high-temperature limit (neglecting static spin correlations) by components of DM vectors in the laboratory frame as[18]

$$M_2^{DM} = \frac{S(S+1)}{3N}\sum_{(ij)}\left((D_{ij}^x)^2 + (D_{ij}^y)^2 + 2(D_{ij}^z)^2\right),$$

$$M_4^{DM} = \frac{S^2(S+1)^2}{N}\left\{\frac{2}{3}\sum_{(ij)}J_{ij}^2\left((D_{ij}^x)^2 + (D_{ij}^y)^2 + 2(D_{ij}^z)^2\right)+ \right.$$

$$\left. + \frac{1}{18}\sum_{(ijk)}\left((F_{ijk}^x)^2 + (F_{ijk}^y)^2 + 2(F_{ijk}^z)^2 + (F_{jki}^x)^2 + (F_{jki}^y)^2 + 2(F_{jki}^z)^2 + (F_{kij}^x)^2 + (F_{kij}^y)^2 + 2(F_{kij}^z)^2\right)\right\} \quad (5)$$

where the three-site term sums over functions $F_{ijk}^\alpha = J_{ij}(D_{ik}^\alpha - D_{jk}^\alpha) + J_{ik}(D_{ij}^\alpha + D_{jk}^\alpha)$. The expected ESR linewidth anisotropy is then obtained from Eqs. (4) and (5) by transforming the DM vectors from the laboratory frame to the crystal frame

$$D_{ij}^x = D_{ij}^a \cos\theta\cos\varphi + D_{ij}^b \cos\theta\sin\varphi - D_{ij}^c \sin\theta,$$

$$D_{ij}^y = -D_{ij}^a \sin\varphi + D_{ij}^b \cos\varphi, \quad (6)$$

$$D_{ij}^z = D_{ij}^a \sin\theta\cos\varphi + D_{ij}^b \sin\theta\sin\varphi + D_{ij}^c \cos\theta.$$

We denote an arbitrary orientation of applied constant magnetic field ($z$-direction in the laboratory frame) with the polar angle $\theta$ and the azimuthal angle $\phi$ in the crystal frame of $SrCu_2(BO_3)_2$ system with **c**-axis as the polar axis.

## IV. Results

Lorentzian shape of very broad X-Band ESR absorption lines measured in a single crystal of $SrCu_2(BO_3)_2$ has already been reported earlier.[19] The lineshape remains Lorentzian even at 580 K proving that spin-diffusional behavior of the spin correlation function is negligible. The temperature dependence of the linewidth shows a broad minimum around room temperature (RT) for both extreme orientations of the external static magnetic field **B₀** with respect to the crystal anisotropy **c**-axis as shown in Fig. 2. The increase of the linewidth below RT is significant in both directions and reflects spin dynamic governing the evolution of the spin correlation function. On the other hand, above RT the linewidth exhibits at first site rather surprising linear behavior with a slope almost independent on the angle $\theta$ between the external field and crystal **c**-axis, i.e. the slope equals 0.355±0.01 G/K for $\theta = 0°$ and 0.325±0.01 G/K for $\theta = 90°$ (inset to Fig. 2). The unusual high-$T$ increase has already been discussed on the qualitative grounds.[19] Its origin was proposed to arise from spin-orbit coupling giving rise to additional line-broadening mechanism. However, in present paper also a quantitative description is given on a basis of a phonon modulation of Dzyaloshinsky-Moriya interaction. We shall come back to this linear contribution to the ESR linewidth latter. For the moment, let us subtract this part from the measured linewidths and focus on the angular dependence.

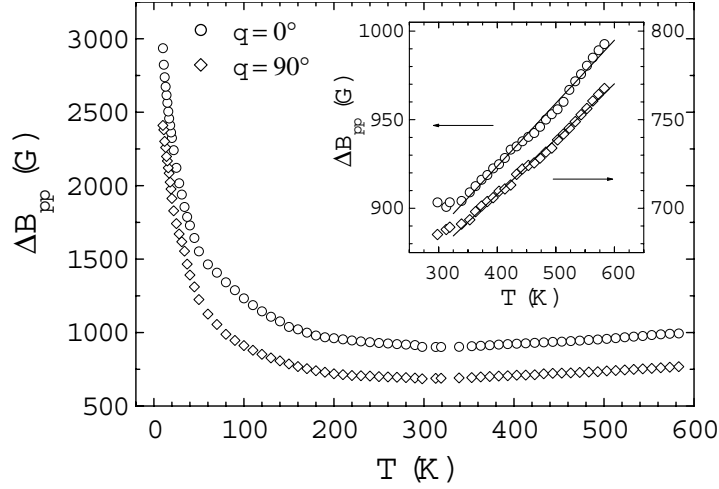

**Fig. 2:** Temperature dependence of the linewidth of the X-band ESR spectra from a SrCu$_2$(BO$_3$)$_2$ single crystal for **B**$_0$||**c** (circles) and **B**$_0\perp$**c** (diamonds). Inset shows the high-$T$ linearly increasing behavior.

Angular dependence of the linewidth at 295 K and 525 K is shown in the inset to Fig. 3. It is significant in the crystal *ac*-plane (dependence on the polar angle $\theta$) while there is almost no angular dependence with respect to the azimuthal angle $\phi$ (*ab*-plane).

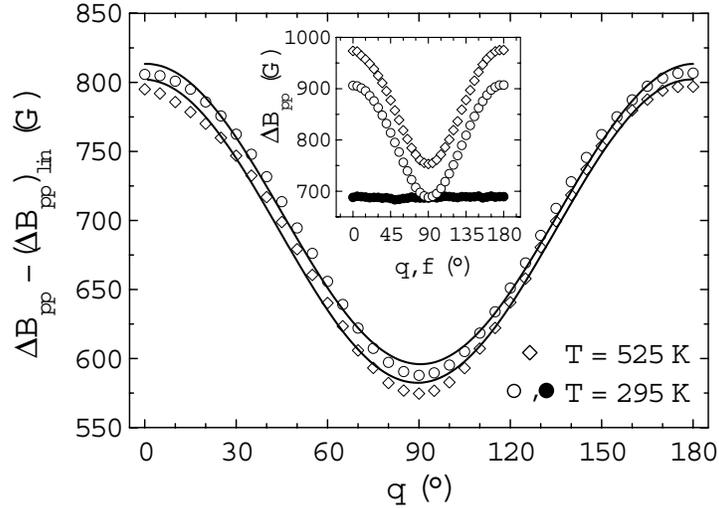

**Fig. 3:** X-band ESR anisotropy of a SrCu$_2$(BO$_3$)$_2$ single crystal with the high-$T$ linear part subtracted at 525 K (diamonds) and 295 K (circles). The solid lines correspond to angular dependence predicted by Eq. (8) with $D'_\| = 2.3\,\text{K}$ and $D_\perp = 4.0\,\text{K}$. The inset shows "raw" linewidth anisotropy in the crystal *ac*-plane (open symbols) at 525 K (diamonds) and 295 K (circles) and in the *ab*-plane at 295 K (full circles).

The $\theta$-dependence of the linewidth at both temperatures can be qualitatively described by an equation of the form $A + B(1 + \cos^2\theta)$. We use such a form rather than $A' + B'\cos^2\theta$ because of the characteristic linewidth anisotropy expected for DM interaction as it will be revealed in the next chapter. The experimentally determined values of parameters are $A$ = 378±5 G, $B$ = 218±5 G for $T$ = 295 K and $A$ =362±5 G, $B$ =220±5 G for $T$ = 525 K when the high-temperature part is excluded. As the high-$T$ increasing part of the linewidth has more or less the same slope for $\theta$ = 0° and $\theta$ = 90° the same set of parameters $A$ and $B$ corresponds to the linewidth anisotropy for temperatures above approximately 340 K (see inset to Fig. 2).

Due to the large exchange coupling only few spin-anisotropy interactions are expected to account for experimental linewidths. Actually, major anisotropy contributions to the linewidth have already been evaluated in our previous report,[19] where Dzyaloshinsky-Moriya interaction was shown to play by far the biggest role. In fact, among the major magnetic anisotropy interactions expected for a spin S = 1/2 system, namely, the dipolar interaction, the hyperfine coupling, the symmetric anisotropic exchange and the antisymmetric DM interaction, the latter interaction is the only one giving linewidths of the correct order of magnitude.

## V. Discussion

### A. Linewidth anisotropy

Exploring the proposed picture of only interdimer DM interactions with corresponding DM vectors parallel to crystal **c**-axis proposed by Cépas et al.,[10] the angular dependence of the linewidth can be derived from Eqs. (4), (5) and (6)

$$\Delta B_{pp} \approx \frac{2\pi}{\sqrt{6}} \frac{k_B^2}{g\mu_B} \frac{D_\parallel'^2}{\sqrt{6(3J^2 + 3J'^2 - 2JJ')}} \left(1 + \cos^2\theta\right). \tag{7}$$

The linewidth is expected to possess no $\phi$-dependence, which is in an excellent agreement with the experiment (Fig. 3). On the other hand, the $\theta$-dependence of the form $B(1+\cos^2\theta)$ is predicted. However, the experimental results presented in Fig. 3 differ from this prediction in one important detail. Namely, there is an additional constant parameter $A$, which has to be included to yield a satisfactory fit. We stress out once again that the high-temperature linearly increasing part is subtracted from the "raw" data. As the parameter $A$ is approximately 1.6-times larger than parameter $B$, which is responsible for the observed linewidth anisotropy, the magnetic anisotropy introducing the additional $A$ term should be substantial and cannot be simply ignored. This raises a fundamental question about the origin of this supplementary contribution to the linewidth. Symmetric anisotropy interaction, which should give the second largest contribution to the linewidth, can be safely dismissed from the possible causes of the observed discrepancy for at least two reasons. It is more than an order of magnitude to small[19] and it results in an "incorrect" angular dependence. To be precise, the symmetric anisotropic exchange interaction can be presented by a traceless tensor with its principal axes coinciding with those of the $g$-tensor, which is in the case of the investigated compound axially symmetric to a very good approximation.[19] This yields an angular dependence again of the form[20] $(1+\cos^2\theta)$, thus failing to give a satisfactory justification of the additional *A* term.

We have already proposed in our initial report[19] that the observed angular dependence of the linewidth originates from additional magnetic-anisotropy terms of Dzyaloshinsky-Moriya type that should be included to the spin Hamiltonian. We now proceed with a model involving also an intradimer DM interaction and prove that this interaction is needed to understand the angular and temperature dependence of the X-band ESR linewidth in SrCu$_2$(BO$_3$)$_2$. Additionally, we propose a picture describing the nature of intradimer DM term based on minor distortions of the crystal-structure of investigated compound, which leads to a consistent explanation of the observed linewidth anisotropy for all temperatures above RT.

As already emphasized, when treating the CuBO$_3$ crystal planes as ideally planar only the interdimer DM interactions with vectors perpendicular to this plane are allowed for symmetry reasons. On the other hand, a finite buckling of these planes has been reported below the displacive, second order structural phase transition occurring at $T_s$ = 395 K.[12] Bending of neighboring CuO$_4$ plaquettes, which evolves progressively below the phase transition temperature $T_S$, reaches a bending angle of 8° at RT and 11° at 100 K. Such distortion loosens the symmetry restrictions placed on the DM interaction, as the crystal *ab*-plane is not a mirror plane anymore. This corrugation of CuBO$_3$ planes allows for finite in-plane components of interdimer DM interaction $D'_\perp$ as well as for in-plane intradimer coupling $D_\perp$. Moreover, the direction of DM vectors of the latter interaction is well defined due to the fact that two mirror planes perpendicular to crystal *ab*-plane, one of them containing a dimer and the other one perpendicular to it, still exist. Intradimer DM vectors thus have to point perpendicularly to the direction of the dimer (Fig. 1). The interdimer $D'_\perp$ coupling has recently been proposed to be responsible for the fine structure of the single triplet excitation at wave vector $Q$ = (1.5,0,0) observed in high-resolution inelastic neutron scattering experiments.[21] This coupling has been estimated to be of the order of 30% of $D'_\parallel$ in the $J'$ = 0 limit. Since the ESR linewidth is proportional to the square of the amplitude of DM vector, the in-plane interdimer component of DM interaction yields smaller corrections to the linewidth. For this reason it will not be considered in the first approximation.

Taking into account also the intradimer DM interaction $D_\perp$ in addition to the well-approved interdimer one $D'_\parallel$ the expression of the ESR linewidth (Eq. (4)) together with the expressions of the second and the forth moment (Eqs. (5) and (6)) lead to the following linewidth anisotropy

$$\Delta B_{pp} \approx \frac{2\pi}{\sqrt{6}} \frac{k_B^2}{g\mu_B} \left( \frac{\left(8D'^2_\parallel + 3D_\perp^2 + \left(8D'^2_\parallel - D_\perp^2\right)\cos^2\theta\right)^3}{96\left(32D'^2_\parallel J_1^2 + 3D_\perp^2 J_2^2 + \left(32D'^2_\parallel J_1^2 - D_\perp^2 J_2^2\right)\cos^2\theta\right)} \right)^{1/2}, \quad (8)$$

with $J_1^2 = 3J^2 + 3J'^2 - 2JJ'$ and $J_2^2 = 13J^2 + 6J'^2$. Due to the fact that dimers form orthogonal network the linewidth remains $\phi$-independent, which is, as already stressed, consistent with the experiment. The derived equation (8) fits the observed linewidth anisotropy rather well as shown in Fig. 3. It should be emphasized that Eq. (8) consistently describes the data at all temperatures and that no additional term is needed once the high-temperature linearly increasing part is eliminated. The parameters obtained for 525 K are $D'_\| = 2.4 \pm 0.1\,\text{K}$ and $D_\perp = 4.0 \pm 0.1\,\text{K}$. At 295 K the parameters are only slightly bigger, i.e. by approximately 0.1 K as expected from Fig. 3. Moreover, the high-$T$ parameters seem to be valid for all temperatures above approximately 340 K as the linewidth simply increases with similar slope for all directions above this temperature. The $D'_\| = 2.4\,\text{K}$ term is in good agreement to the previous estimation of the interdimer DM interaction.[10] It further justifies the use of the constant $C = 2\pi/\sqrt{6}$ in Eq. 4 characteristic for a lineshape, which is determined by a product of Lorentzian line with a Gaussian line. On the other hand, the value of the nearest-neighbor DM interaction $D_\perp = 4.0\,\text{K}$ seems to be unexpectedly high at first glance. Yet, it is consistent with recent high-field ESR measurements,[22] where the authors suggest that in-plane intradimer DM interaction should be substantial based on relative intensities of observed singlet-triplet transitions.

We also considered the possibility of a finite in-plane interdimer DM interaction (Fig. 1). Let the DM interaction between *nnn* spins 1 and 3 be of the form $\mathbf{D}_{13} = (D'_\xi, D'_\eta, D'_\|)$. The four symmetry operators of the SrCu$_2$(BO$_3$)$_2$ space group ($I\bar{4}2m$) below $T_s$ = 395 K, namely, the mirror planes $m_1\; x,x,z$, $m_2\; x,\bar{x},z$, and the fourfold rotoinversion axes $\bar{4}^-\; 0,1/2,z; 0,1/2,1/4$, $\bar{4}^+\; 1/2,0,z; 1/2,0,1/4$, then determine all the remaining *nnn* DM vectors as indicated in Fig. 1. With this information the expression for the linewidth can be derived in a similar way as Eqs. (7) and (8) have been obtained. The result is a coupling of in-plane intradimer and interdimer DM components while the out-of-plane interdimer component remains unchanged $D'_\|$. Thus, from ESR linewidth anisotropy one cannot distinguish between the in-plane *nn* and *nnn* DM vectors. Taking the later interaction from high-resolution inelastic neutron scattering results[21] to amount for $D'^2_\xi + D'^2_\eta = D'^2_\perp \approx 0.1 D'^2_\|$ our fit gives smaller value of the former interaction at 525 K, i.e. $D_\perp = 3.6\,\text{K}$. However, since the observed anomalous splitting of the triplet excitation at $Q$ = (1.5,0,0) is not necessary solely due to the in-plane interdimer DM component, there is certain ambiguity about the size of $D_\perp'$. At this stage we can thus only make an estimation that the intradimer DM interaction is of the order of $D_\perp = 3.6 \pm 0.5\,\text{K}$.

The temperature evolution of the linewidth anisotropy above RT poses a question about the nature of the in-plane DM components employed in the above analysis since the buckling of CuBO$_3$ planes progressively disappears when approaching the structural phrase transition temperature of $T_s$ = 395 K from below. Consequently, above this temperature only interdimer DM vectors parallel to crystal **c**-axis are allowed in the static lattice picture predicting (1+cos$^2\theta$) angular dependence. However, from the present experiment (Fig. 2) it seems that the linewidth anisotropy is not affected by this structural phase transition at all. Trying to give a credible explanation for this puzzle, one is forced to employ also dynamical effects. In fact, dynamical Dzyaloshinsky-Moriya interaction has already been proposed as a possible mechanism for the explanation of the singlet-triplet transitions in high-field ESR experiments.[23] In this picture lattice vibrations instantaneously break local symmetry allowing for additional terms of DM interaction. However, there are at least two requirements to be fulfilled before including this mechanism into the interpretation of our X-band ESR results. First of all, if the intradimer DM coupling were of dynamical origin one would expect to observe its effects when the characteristic phonon frequency is small compared to characteristic exchange frequency $\omega_e \sim k_B J/\hbar$ determining the correlation time $\tau_c$ of the decay of spin correlation function $\varphi(t)$. Since optical phonons are needed to produce required lattice distortions, significant softening of a particular normal mode should be present. Secondly, the mean square displacements of ions, participating in the aforementioned lattice motion, from their equilibrium positions should be large enough to break the local symmetry significantly, i.e. more than it is already broken by finite buckling at a particular temperature if present. It seems that both conditions are fulfilled in $SrCu_2(BO_3)_2$ system. Many structural properties of $SrCu_2(BO_3)_2$ system have, fortunately, been carefully investigated by Sparta et al.[12] Of our particular interest is their observation of a soft mode detected by Raman scattering experiment. This particular normal mode has been reexamined in a very recent study combining new Raman scattering results with the powerful symmetry analysis of normal modes is $SrCu_2(BO_3)_2$ system.[24] The Raman shift of this optical mode amounts to 62 cm$^{-1}$ (89 K) at $T$ = 15 K, which is very close to the value of the intradimer exchange coupling constant. Moreover, it progressively softens by 44 cm$^{-1}$ with increasing temperature before disappearing into the tail of quasielastic scattering around the phase transition temperature $T_s$ = 395 K. Its zone-center symmetry $A_1$ corresponds to in-phase motion of almost all ions within the primitive cell (with exception of $Sr^{2+}$ ions) preferably along the crystal **c**-direction. The evolution of this mode follows subgroup connections between the space groups of the low-temperature and the high-temperature phase when crossing the structural phase transition temperature, which allowed the authors to assign one of the four Raman silent $B_{1u}$ buckling modes of the high-temperature phase as a soft mode of this phase. In fact, below $T_s$ some in-phase superposition of Cu, B and the two non-equivalent oxygen (O1 and O2) normal coordinates becomes frozen-in, which causes finite buckling. The symmetry of this soft mode is also consistent with the observed anomalous anharmonicity of lattice properties as revealed by X-ray diffraction measurements.[12] The anharmonicity is reflected in a flattening of the local potentials of Cu, B, O1 and O2 ions as well as in a significant enhancement of their mean square displacements, both phenomena occurring only in the crystal **c**-direction and progressively getting larger when approaching the phase transition temperature from below. Thus it seem that near the phase transition temperature and especially above this temperature soft phonon modes are active with ions vibrating in-phase in the crystal **c**-direction with very low frequencies and at large amplitudes. For instance the $U_{33}$ principal components of the displacement tensor, shown in the report for Cu and O1 ions,[12] yield enhanced vibration amplitudes in crystal **c**-direction in the high-temperature phase, which are virtually the same as static displacements (0.25 Å for Cu ions and 0.33 Å for O1 ions) of these ions in the low-temperature phase at temperatures as low as 100 K. From X-band ESR point of view such lattice vibrations break the local symmetry in a very similar way as it is broken in the low-temperature phase by finite buckling since their frequency is very small compared to the exchange frequency $\omega_e$. Another proposal from ref. 12 worth mentioning is that in-plane exchange coupling constants $J$ and $J'$ are not expected to vary considerably with temperature despite the corrugation of $CuBO_3$ planes. Therefore, it is not surprising that the ESR linewidth anisotropy with the linear part subtracted does not change in as broad region as from RT to 580 K despite crossing the structural phase transition at $T_s$ = 395 K. We propose that either the low temperature source of the intradimer DM interaction, if of static origin (i.e. buckling of CuBO$_3$ planes), evolves progressively into a dynamical source with increasing temperature or the source is of dynamical origin in the whole temperature range above RT as the vibration amplitudes are sizable even at 100 K (0.1 Å).[12]

By introducing additional anisotropic terms of DM type to the spin Hamiltonian of the system, one has to consider carefully whether the observed fine structure of triplet excitations can be theoretically reproduced. In fact, it has been shown using the finite temperature Lanczos (FTL) method on a finite-size Shastry-Sutherland lattice that the effect of intradimer DM interaction on the fine structure is only minor at low magnetic fields so that the splitting remains primarily given by the out-of-plane interdimer DM interaction.[25]

The estimation of the intradimer DM interaction with a vector **D** = (2.2K, 2.2K, 5.2K) has been reported quite recently.[26] The inclusion of such nearest-neighbor DM coupling is needed to reproduce the temperature dependence of the specific heat at low temperatures and in high magnetic fields (above 27 T). The report yields in-plane intradimer interaction of a size 3.1 K, which is comparable but somewhat smaller than our estimation. However, there is also a sizable out-of-plane component of this interaction, which should be constrained to a value of zero by the crystal symmetry. The authors argue that such DM vectors could be due to a lattice distortion induced by magnetic field as evidenced by ultrasonic experiments.[27] In this sense, their report cannot be simply compared to the estimations obtained in the current paper, especially, since intradimer DM coupling has no effect on *T*-dependence of the specific heat in low magnetic fields.[28]

## B. High-temperature linewidth behavior

Next we focus on the temperature dependence of the linewidth, in particular on the high-temperature linearly increasing part. For a pure spin Hamiltonian ESR linewidths are normally expected to approach a constant value for temperatures $T \gg J$, when the difference of Boltzman population factors is safely neglected. However, spin-diffusional decay of spin correlation function may become significant at high temperatures leading to temperature dependent ESR linewidth.[17] Since the ESR lineshape in SrCu$_2$(BO$_3$)$_2$ system remains Lorentzian even at 580 K this mechanism can be safely ruled out as a possible source of line broadening.[19] If spin diffusion is present in a 2D magnetic systems one expects to observe a lineshape in-between Lorentzian and Gaussian.[17] Secondly, the effects of static spin correlations (i.e. short-range order) on the linewidth can also be observed in low-dimensional magnetic systems even at temperatures of the order of $T \sim 10J$. For an ordinary *nn* square lattice it was shown that static spin correlations lead to temperature dependent $M_2$ of the form[29] $M_2(T,J) = \left(M_2^S F^S(T,J) + M_2^A F^A(T,J)\right) \chi_C / \chi(T)$, where a factorization into angular-dependent part $M_2^{S,A}$ and temperature-dependent part is possible separately for symmetric and antisymmetric contributions to the second moment $M_2^{S,A}(T,J)$. The temperature dependence due to static spin correlations is hidden in the ratio between Curie susceptibility $\chi_c$ and the measured susceptibility $\chi(T)$ as well as in functions $F^{S,A}(T)$ reflecting the temperature evolution of two-site static spin correlations $C_{ij}(T)$ between spins at sites $i$ and $j$. Following the original paper of Soos et al.,[30] the temperature-dependent second moment due to the DM interaction can be calculated

$$M_2^{DM} = \frac{S(S+1)}{3N}\left\{\sum_{(ij)}\left(\left(D_{ij}^x\right)^2 + \left(D_{ij}^y\right)^2 + 2\left(D_{ij}^z\right)^2\right)(1-C_{ij}) + \sum_{\substack{(ij)\\k\neq i,j}}\left(D_{ij}^x D_{ik}^x + D_{ij}^y D_{ik}^y + 2D_{ij}^z D_{ik}^z\right)C_{jk} + \right.$$
$$\left. + \sum_{(ij)}\sum_{\substack{(klj)\\k,l\neq i,j}}\left(D_{ij}^x D_{kl}^x + D_{ij}^y D_{kl}^y + 2D_{ij}^z D_{kl}^z\right)\left(C_{ik}C_{jl} - C_{il}C_{jk}\right)\right\}\frac{\chi_C}{\chi(T)}. \quad (9)$$

In the situation when the temperature-dependent second moment can be factorized into the infinite-temperature second moment times a temperature-dependent part as for an ordinary *nn* square lattice, the change of the linewidth due to short-range correlation effects in a particular temperature interval has the same angular dependence as the infinite-temperature second moment. However, in the investigated system such factorization is not possible and each of the three sums in Eq. (9) yields its own characteristic temperature behavior, with the first one approaching the infinite- temperature second moment and the last two going to zero when increasing the temperature. It is, though, highly unlikely that the two-site correlation function $C_{ij}(T)$ would evolve with temperature in such a way to produce angular-independent line broadening as observed in current investigation. Another reason for dismissing static spin correlations as the origin of the observed linewidth behavior at high temperatures, is almost ideal linear dependence in a rather broad temperature range, i.e. between $3.5J$ and $7J$. In such a broad region one would expect significant bending of the linewidth curves towards the infinite temperature value if the broadening was due to short-range order effects.

We propose that the observed high-temperature increase is due to the significant spin-phonon coupling present in the system. Such a coupling has been observed from the dramatic softening of elastic constants both with temperature and with applied magnetic field.[31] Spin-phonon interaction causes lifetime broadening effects. As the increase of the linewidth is linear, normal direct phonon processes should be involved. In particular, the observed line broadening can be either due to the usual spin-lattice relaxation between Zeeman split triplet levels or a reflection of transitions between the ground state and triplet states induced by a phonon modulation of antisymmetric Dzyaloshinsky-Moriya interaction.[32] For the former mechanism a strong field-dependent behavior is typical,[33] while no field dependence is expected in the later case.[32] High-field ESR spectra of $SrCu_2(BO_3)_2$ system show no additional line broadening with respect to X-band ones. In the inset to Fig. 4 a high-field spectrum of a powdered sample, recorded at a frequency of 93.2 GHz at RT is shown. As it can be seen, it is nicely fitted with a Lorentzian function for powder spectra yielding anisotropic linewidths of $\Delta B_{pp}^a = \Delta B_{pp}^b = 699\pm 10\,\text{G}$ and $\Delta B_{pp}^c = 890\pm 10\,\text{G}$. These linewidths are virtually the same as the single-crystal X-band values $\Delta B_{pp}^{\theta=90°} = 690\pm 5\,\text{G}$ and $\Delta B_{pp}^{\theta=0°} = 907\pm 5\,\text{G}$. Since the linear contribution to the linewidth at RT is of the order of 100 G and is the same at two resonance fields differing by almost a factor of 10, we propose that the spin-lattice contribution to the linewidth is due to fluctuating DM interaction.

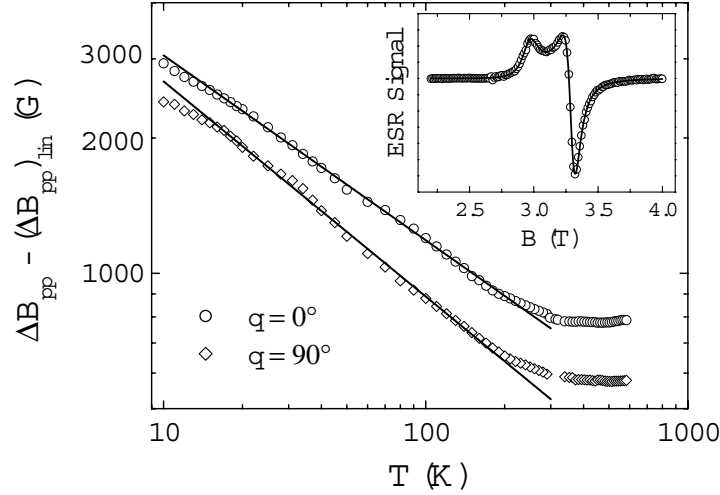

**Fig. 4:** Log-to-log plot of the linewidth of X-band ESR spectra from a $SrCu_2(BO_3)_2$ single crystal with the high-$T$ contribution subtracted for $\mathbf{B}_0 \parallel \mathbf{c}$ (circles) and $\mathbf{B}_0 \perp \mathbf{c}$ (diamonds). Full lines represent fits to an equation of the form $\Delta B = A/(T-T^*)^p$. The inset shows a high-field ESR spectrum of a powdered $SrCu_2(BO_3)_2$ sample at RT (circles) and a fit to anisotropic Lorentzian powder distribution (full line).

Making a rather crude approximation with neglecting correlation effects between a pair of interacting spins and their neighbors, the linewidth is determined by phonon-induced transitions between a singlet state and triplet states of a two spin system. Since $J$ is two orders of magnitude larger than characteristic Zeeman energy and the phonon density scales as $\omega^4$, it is not surprising for the modulated-DM interaction effect to dominate over phonon-induced Zeeman transition between triplet states. In a simple picture of uncorrelated dimers the finite-lifetime contribution to the linewidth is given by equation18

$$\Delta B_{pp}^{lin} \approx \frac{8z}{9\sqrt{3}} \frac{k_B^4}{g\mu_B} \frac{(\lambda R)^2 D^2 J^2}{\rho \hbar^3} \left\langle \frac{1}{c_t^5} + \frac{2}{3}\frac{1}{c_l^5} \right\rangle_\Omega k_B T . \qquad (10)$$

Taking the number of independent pairs as $z = 1$, $\lambda R = 10$ with the nearest-neighbor distance $R$ and $dJ/dr = -\lambda J$,[34] $\rho = 4.1$ kg/dm$^3$,[27] $D = D_\perp = 3.6$ K and an approximate average velocity[31] $c = 4600$ m/s in place of the complicated angular average in Eq. (10), we can estimate the slope of the linearly-increasing part to be of the order of 0.14 G/K. This result is in a reasonable agreement with the experimentally observed slope 0.34 G/K bearing in mind that the mean velocity, which is burdened with the biggest uncertainty, is taken to the power of 5. Moreover, Eq. (10) is strictly valid only in a crude approximation of independent dimers. Since interactions with other neighbors in general shorten the lifetime of a spin in a certain energy level, the slope is expected to be larger in the real system than the value our estimation produces.

## C. Low-temperature linewidth behavior

In the end let us give a brief discussion on the temperature dependence of the ESR linewidth below RT. Fig. 4 shows a log-log plot of the linewidth without the high-temperature linear part. It should be emphasized that the high-temperature contribution is expected to deviate from linear dependence at low temperatures since the mean excitation number of phonons with energy $k_B J$ changes as $n = (\exp(-J/T) - 1)^{-1}$. However, the spin-phonon contribution to the linewidth at low-temperatures becomes insignificant, which hardly makes any difference when a linear contribution or a contribution proportional to the actual number of phonons, $n$, is subtracted from the linewidth.

The evolution of the low-temperature part of the linewidth with temperature yields a linear dependence in the log-log plot to a reasonably good approximation in a rather broad temperature range between 15 K and 150 K. Solid lines correspond to a fit to the function $\Delta B = A/(T - T^*)^p$, characteristic of critical broadening effects. The obtained values for the exponent are $p = 0.41 \pm 0.01$ for direction parallel to crystal **c**-axis and $p = 0.47 \pm 0.01$ for $\theta = 90°$, while the value of the characteristic temperature $T^*$ is within the experimental error very close to zero. We have reported similar observations on the linewidth behavior in powder samples.[15] Such critical behavior typically occurs near phase-transition temperatures with induced long-range order in the low-temperature phase. Its presence, if confirmed in future work, would be in the case of the investigated system probably due to critical enhancement of antiferromagnetic fluctuations at wave-vector at the zone boundary of the first Brillouin zone. A manifestation of this mechanism in ESR linewidths would be an unambiguous fingerprint of the importance of quantum criticality effects in the $SrCu_2(BO_3)_2$ system, as the system is believed to be situated in the extreme vicinity of the quantum borderline between the non-magnetic singlet ground state and magnetically ordered ground state. However, the observed temperature interval (15 K to 150 K), in which this theory seems to be consistent, is surprisingly broad. In this respect, the results of our investigation call for a comprehensive theoretical evaluation of the temperature evolution of the relaxation function, reflecting spin dynamics present in the $SrCu_2(BO_3)_2$ system. Such theory would also have to account for the observed small deviations from "linear" dependence below 15 K.

## VI. Conclusions

In summary, a comprehensive X-band ESR study on a single crystal of highly frustrated $SrCu_2(BO_3)_2$ system has been given in a broad temperature range between 10 K and 580 K. The rather broad absorption lines were consistently explained by the antisymmetric Dzyaloshinsky-Moriya interaction, which yields the largest contribution to the second and the forth moment of ESR spectra. The anisotropy of the linewidth shed some important additional light on the spin anisotropy of the investigated systems. Previously proposed out-of-plane interdimer DM interaction as the dominant anisotropic contribution, cannot adequately account for the experimental angular dependence. In fact, one has to incorporate also sizable in-plane DM components in addition to the out-of-plane interdimer coupling $D'_\parallel = 2.4 \pm 0.1\,\text{K}$. Taking the value of the in-plane interdimer DM vector $D'_\perp$ to be 30% of the out-of-plane one, as suggested by recent inelastic neutron scattering experiments, allowed us to evaluate also the intradimer DM interaction, $D_\perp = 3.6 \pm 0.5\,\text{K}$. Moreover, as in the high-temperature structural phase (above 395 K) the in-plane components of DM interaction are prohibited by symmetry, an explanation of the nature of these terms based on a dynamical picture including a soft mode has been given. The observed highly anisotropic linewidth shows a peculiar temperature dependence, which can be decomposed into two contributions. The high-temperature contribution is arising from the interplay between spin and lattice degrees of freedom, while the low-temperature part is due to spin dynamics present in the spin system. The high-temperature component has been proposed to be related to the induced transitions between energy levels due to phonon-modulation of the Dzyaloshinsky-Moriya interaction. On the other hand, the low-temperature part exhibits critical-broadening-like behavior, which could be a consequence of enhanced antiferromagnetic correlations.

## Acknowledgements

It is a pleasure to thank Dr. A. Lappas for stimulating discussions on the magnetic properties of $SrCu_2(BO_3)_2$. HK thanks the Yamada Science Foundation (Japan) for supporting his work in low-dimensional systems.